\documentclass[aps,prd,12pt,nofootinbib,notitlepage]{revtex4-1}

\usepackage{amsfonts}
\usepackage{amsmath,epsfig,bm}
\usepackage{graphicx}
\usepackage{bm}
\usepackage{color}

\usepackage{tabularx} 

\usepackage{times}

\newcommand{\be}{\begin{equation}}
\newcommand{\ee}{\end{equation}}
\newcommand{\ba}{\begin{eqnarray}}
\newcommand{\ea}{\end{eqnarray}}
\newcommand{\bd}{\begin{displaymath}}
\newcommand{\ed}{\end{displaymath}}

\begin{document}

\title{Perspective on Tsallis Statistics for Nuclear and Particle Physics}
\author{Joseph I. Kapusta}
\affiliation{School of Physics and Astronomy, University of Minnesota, Minneapolis, Minnesota 55455, USA}

\vspace{.3cm}

\begin{abstract}
This is a concise introduction to the topic of nonextensive Tsallis statistics meant especially for those interested in its relation to high energy proton-proton, proton-nucleus, and nucleus-nucleus collisions.  The three types of Tsallis statistics are reviewed.  Only one of them is consistent with the fundamental hypothesis of equilibrium statistical mechanics.  The single particle distributions associated with it, namely Boltzmann, Fermi-Dirac, and Bose-Einstein, are derived.  These are not equilibrium solutions to the conventional Boltzmann transport equation which must be modified in a rather nonintuitive manner for them to be so.  Nevertheless the Boltzmann limit of the Tsallis distribution is extremely efficient in representing a wide variety of single particle distributions in high energy proton-proton, proton-nucleus, and nucleus-nucleus collisions with only three parameters, one of them being the so-called nonextensitivity parameter $q$.  This distribution interpolates between an exponential at low transverse energy, reflecting thermal equilibrium, to a power law at high transverse energy, reflecting the asymptotic freedom of QCD.  It should not be viewed as a fundamental new parameter representing nonextensive behavior in these collisions.
\end{abstract}

\maketitle

\section{Introduction}

In 1988 Constantino Tsallis \cite{Tsallis1988} entertained the possibility of ensembles where the entropy took a nonadditive form involving a parameter $q$ which reduced to the usual entropy in the limit $q = 1$. Since then the so-called Tsallis statistics has been applied to many areas of the natural and social sciences \cite{Tsallisbook}.  There are approximately 9,000 citations to this paper according to Google Scholar, and 1,500 articles in the preprint arXiv with ``Tsallis" in the abstract.  In this perspective I give a concise introduction to and interpretation of how Tsallis statistics has been used in high energy proton-proton, proton-nucleus, and nucleus-nucleus collisions.

Although commonly used, ``Tsallis statistics" is a bit of a misnomer.  It is usually associated with an energy distribution, to be discussed more thoroughly below.  Quantum field theory allows for two types of particle statistics: Bose and Fermi.  An arbitrary number of bosons can occupy a quantum state whereas at most one fermion can.  In the Boltzmann limit the occupation probability is much smaller than 1 so that it does not matter whether the particle is a boson or a fermion.  

Ludwig Boltzmann defined the entropy in statistical mechanics as the logarithm of the number of available states $\Omega$
\be
S = \ln \Omega
\label{BS}
\ee
(For notational convenience we use units with $\hbar = c = k_B = 1$.)  J. Willard Gibbs defined it as
\be
S = - \sum_i p_i \ln p_i
\label{GS}
\ee
where $i$ labels an eigenstate of the Hamiltonian $H$, and $p_i$ is the probability of that state being occupied within the ensemble.  John von Neumann defined it as
\be
S = - {\rm Tr} (\rho \ln \rho)
\label{vNS}
\ee
where $\rho$ is the statistical density matrix.  It reduces to Eq. (\ref{GS}) in a diagonal basis.  Claude Shannon defined an information entropy the same as Eq. (\ref{GS}).  For independent systems $A$ and $B$ the number of states is the product $\Omega = \Omega_A \Omega_B$ and so $S = S_A + S_B$ is additive, as it ought to be in macroscopic thermodynamics.  If the two systems are in physical contact then this ignores surface terms compared to volume terms.  The same reasoning applies when the entropy is expressed in terms of probabilities since the probabilities for independent systems multiply.

Tsallis considered the possibility of systems where the entropy instead had the form
\be
S = \frac{1}{q-1} \sum_i (p_i - p_i^q)
\label{Tentropy}
\ee
where $q$ is a real parameter.  This has the property that $S(q \rightarrow 1)$ yields the usual expression for the entropy.  If there are two independent systems $A$ and $B$ such that the probabilities multiply $p_i \rightarrow p_{Ai} p_{Bj}$ and the probabilities are normalized 
\be
\sum_i p_{Ai} = \sum_i p_{Aj} = 1
\ee
then the total entropy is not the sum of the entropies of each system.  Instead
\be
S_{A+B} = S_A + S_B + (1-q) S_A S_B
\ee
meaning that the entropy as defined is not an additive quantity.  Similarly for three independent systems
\be
S_{A+B+C} = S_A + S_B + S_C + (1-q) (S_A S_B + S_A S_C + S_B S_C) + (1-q)^2 S_A S_B S_C
\ee
Hence $q$ is referred to as the nonextensivity parameter.  It is said that the Tsallis entropy (\ref{Tentropy}) can represent some physical systems with a long ranged force so that systems $A$ and $B$ are not truly independent.  For example, Ref. \cite{Cirto2018} studied a classical XY model in $d$ dimensions with a 2-body potential that varied as $1/r^{\alpha}$ using molecular dynamics simulations.  For long ranged potentials with $\alpha/d < 1$ it was found that, with the given initial conditions, the energy distribution relaxed to the Tsallis form (see below), whereas for shorter ranged potentials it relaxed to the Boltzmann one.  Reference \cite{Carati2020} studied a realistic ionic crystal model in $d=3$ which includes a long range Coulomb potential.  With two choices of initial conditions, the energy distribution evolves into the Tsallis form at intermediate times albeit with $q(t)$ decreasing with time.  At late times it relaxes to the Boltzmann one with 
$q \rightarrow 1$.  


The average of an operator must also be defined to complete the statistics of the ensemble.  Tsallis and collaborators proposed three \cite{Tsallis1998}.  They and some variants are described in Sec. II.  The Tsallis entropy is used together with the microcanonical ensemble to determine which of the forms are consistent with the fundamental hypothesis of equilibrium statistical mechanics in Sec. III.  The resulting single particle distributions for fermions, bosons, and classical particles are determined in Sec. IV.  A postulated Boltzmann-type transport equation which has the Tsallis distribution as an equilibrium solution is described in Sec. V.  Also discussed in that section is a Fokker-Planck equation which gives rise to a steady-state solution consistent with the Tsallis distribution in certain limits.  High energy collision phenomenology is summarized in Sec. VI.  Section VII contains the conclusions.  I have consistently used $E$ to represent the total energy of a system while $\omega$ represents a single-particle energy.

\section{Tsallis Types}

Tsallis and collaborators proposed three types of operator averages to go along with the entropy Eq. (\ref{Tentropy}).  In this section I describe each of them which are numbered in chronological order.  There are in principle an infinite variety of variants, one of which is also described, along with some observations and comments on all of them.

\subsection{Tsallis Type I}

Type I assumes that probabilities are normalized in the usual way
\be
\sum_i p_i =1
\ee
as are averages of an operator ${\cal O}$ that commutes with $H$
\be
\langle {\cal O} \rangle = \sum_i p_i {\cal O}_i
\ee
where ${\cal O}_i = \langle i | {\cal O} | i \rangle$.  The $|i\rangle$ are eigenstates of $H$ such that $H |i\rangle = E_i |i\rangle$.  The probabilities are found by maximizing the entropy at fixed energy $E = \langle H \rangle$ subject to the normalization condition.
\be
\frac{\delta}{\delta p_k} \left[ S - \beta E - \alpha \sum_i p_i \right] = 0
\ee
Here $\alpha$ and $\beta$ are Lagrange multipliers; $\alpha$ is adjusted to normalize the probabilities and $\beta$ is adjusted to fix the energy.  The resulting distribution is
\be
p_k = \left[ \frac{1 + \alpha (1-q)}{q}\right]^{\frac{1}{q-1}}
\left[ 1 + \frac{(1-q) \beta E_k}{1 + \alpha (1-q)} \right]^{\frac{1}{q-1}}
\ee
There is an upper limit on the allowed values of energy $E_k$ if $q > 1$ and a lower limit if $q < 1$.  This means that some states may not be allowed in the ensemble, which is rather peculiar.  The distribution is not invariant under an overall shift in energy.  Normally, apart from gravity, only energy differences matter.  There are situations where divergences with this type of distribution can occur \cite{Tsallis1998}.  For independent systems $A$ and $B$ the energy is additive
\be
E_{A+B} = E_A + E_B
\ee
even though the entropy is not.

\subsection{Tsallis Type II}

Type II assumes that probabilities are normalized in the usual way
\be
\sum_i p_i =1
\ee
but that averages of an operator ${\cal O}$ are computed in an unconventional way
\be
\langle {\cal O} \rangle = \sum_i p_i^q {\cal O}_i
\ee
This leads to the probability distribution
\be
p_k = \left[ \frac{1 + \alpha (1-q)}{q}\right]^{\frac{1}{q-1}}
\left[ 1 + (q-1) \beta E_k \right]^{-\frac{1}{q-1}}
\ee
There is a lower limit on the allowed values of energy $E_k$ if $q > 1$ and an upper limit if $q < 1$.  As with Type I some states may not be allowed in the ensemble.
The distribution is not invariant under an overall shift in energy.  The divergences that sometimes occur with the Type I distribution do not occur, but this choice has the strange feature that $\langle 1 \rangle \neq 1$ in general.  For independent systems $A$ and $B$ the energy is not additive but satisfies
\be
E_{A+B} = E_A + E_B + (1-q) (S_A E_B + S_B E_A)
\ee
This is rather odd since the assumption was that $H = H_A + H_B$ with no interaction between the two systems.  A redefinition of the entropy should not introduce interactions between the systems. 

\subsection{Tsallis Type III}

Type III assumes that probabilities are normalized in the usual way
\be
\sum_i p_i =1
\ee
but that averages of an operator ${\cal O}$ are computed in another unconventional way
\be
\langle {\cal O} \rangle = \frac{\sum_i p_i^q {\cal O}_i}{\sum_j p_j^q}
\ee
This leads to the probability distribution
\be
p_k = \left[ \frac{1 + \alpha (1-q)}{q}\right]^{\frac{1}{q-1}}
\left[ 1 + \frac{(q-1) \beta (E_k - E)}{1 + (1-q)S} \right]^{-\frac{1}{q-1}}
\ee
This distribution is invariant under an overall shift in energy; only energy differences matter.  Furthermore the energies of two independent systems add, namely
\be
E_{A+B} = E_A + E_B
\ee
Note that the entropy $S$ and average energy $E$ enter explicitly and nontrivialy in the probability distribution.

\subsection{Variants}

Arguing from the point of view of information theory on incomplete probability distributions,  Q. A. Wang \cite{Wang2001} suggested modifying Type II such that
\be
\sum_i p_i^q =1
\ee
This can be rewritten by defining $w_i = p_i^q$ and $q'=1/q$ so that
\ba
\sum_i w_i &=&1 \nonumber \\
\langle {\cal O} \rangle &=& \sum_i w_i {\cal O}_i \nonumber \\
S &=& \frac{q'}{q'-1} \sum_i (w_i - w_i^{q'})
\ea
This is Type I apart from multiplication of the entropy by $q'$.  In fact one can multiply the entropy (\ref{Tentropy}) by any function $g(q)$ with the property that 
$g(q \rightarrow 1) = 1$ and the limit $q \rightarrow 1$ will still be (\ref{GS}).  For Type I the probability distribution becomes
\be
p_k = \left[ \frac{g(q) + \alpha (1-q)}{g(q) q}\right]^{\frac{1}{q-1}}
\left[ 1 + \frac{(1-q) \beta E_k}{g(q) + \alpha (1-q)} \right]^{\frac{1}{q-1}}
\ee
and similarly for the other types. 

\subsection{Observations and Comments}

A notable property of all three types, including the modification of the entropy by $g(q)$ discussed above, is that in the limit $q \rightarrow 1$ the probability becomes the Boltzmann one
\be
p_i = \frac{1}{Z} e^{- \beta E_i}
\ee
with
\be
Z = \sum_j e^{- \beta E_j}
\ee
Keep in mind that $E_i$ is the total energy of the system in the quantum state $|i\rangle$.

All of the discussion so far concerns the canonical ensemble where conserved charges, such as baryon number, electric charge, strangeness, and so on are fixed.  One may introduce Lagrange multipliers, or chemical potentials, for each conserved charge in the same way as $\beta$ is the Lagrange mulitplier for the conserved energy.  The calculation of the probabilities proceeds in a similar manner, but will not be pursued here. 

D. H. Zanette and M. A. Montemurro \cite{Zanette} showed that for any observed probability distribution function $p(X_i)$ of the variable $X$ one can always find a constraint together with the Tsallis entropy which gives that distribution.  For example, if one takes ${\cal O} = \phi(H)$ with Type III then
\be
p(E_k) = \left[ \frac{1 + \alpha (1-q)}{q}\right]^{\frac{1}{q-1}}
\left[ 1 + \frac{(q-1) \beta_{\phi} (\phi(E_k) - \langle \phi \rangle)}{1 + (1-q)S} \right]^{-\frac{1}{q-1}}
\ee 
where $\beta_{\phi}$ is the Lagrange multiplier.  This can be solved for $\phi(E_k)$ in terms of $p(E_k)$.

\section{Microcanonical Ensemble}

The fundamental hypothesis of equilibrium statistical mechanics is that all quantum states with energies between $E$ and $E + \Delta$, with $\Delta \ll E$, are equally likely.  The number of such states is
\be
\Omega(E) = \sum_n \theta (E_n - E) \theta (E + \Delta - E_n)
\ee
and the probabilities for $E_n$ in this range are
\be
p(E_n) = \frac{1}{\Omega(E)}
\ee
and zero otherwise.  This is the microcanonical ensemble.  The standard entropy is
\be
S(E) = \ln \Omega(E)
\ee
whereas the Tsallis entropy is
\be
S(E) = \frac{1}{q-1} \left[ 1 - \left( \frac{1}{\Omega(E)} \right)^{(q-1)} \right]
\ee  
which emphasizes that the difference is in the definition of the entropy.  The temperature associated with the standard definition of entropy is
\be
\frac{1}{T} = \frac{dS}{dE} = \frac{1}{\Omega} \frac{d\Omega}{dE}
\ee
whereas using the Tsallis entropy
\be
\frac{1}{T} = \frac{dS}{dE} = \frac{1}{\Omega^q} \frac{d\Omega}{dE}
\ee

Consider the canonical ensemble which follows from the microcanonical ensemble.  A closed system is divided into parts $A$ and $B$ with total energy $E$.  The probability for part $A$ to be in the state $n$ with energy $E_{An}$ is
\be
p(E_{An}) = \frac{1}{\Omega_{A+B}(E)} \sum_m \theta ( E_{An} + E_{Bm} - E) \theta ( E + \Delta - E_{An} - E_{Bm})
= \frac{\Omega_B (E - E_{An})}{\Omega_{A+B}(E)}
\ee
Let part $A$ be much smaller than part $B$.  Then a Taylor series expansion yields
\ba
S_B(E - \langle E_A \rangle + \langle E_A \rangle - E_{An}) & = &
S_B(E - \langle E_A \rangle) + \frac{d S_B(E - \langle E_A \rangle)}{d (E - \langle E_A \rangle)} \epsilon + \cdot \cdot \cdot\nonumber \\
&=&  S_B(E - \langle E_A \rangle) + \beta \epsilon + \cdot \cdot \cdot
\ea
where $\epsilon = \langle E_A \rangle - E_{An}$ and $\beta$ is the inverse temperature associated with the major part $B$.  Upon exponentiation, and with the usual definition of entropy, this leads to the Boltzmann distribution
\be
p(E_{An}) = \frac{\Omega_B( E - \langle E_A \rangle)}{\Omega_{A+B}(E)} e^{\beta \epsilon} = \frac{1}{Z} \, 
e^{-\beta ( E_{An} - \langle E_A \rangle)}
\ee
whereas with the Tsallis definition of the entropy (\ref{Tentropy}) it leads to
\ba
p(E_{An}) &=& \frac{1}{\Omega_{A+B}(E)} \left[ 1 + (1-q) S_A \right]^{- \frac{1}{q-1}}
\left[ 1 + \frac{(q-1) \beta (E_{An} - \langle E_A \rangle)}{1 + (1-q)S_A} \right]^{-\frac{1}{q-1}} \nonumber \\
&=& \frac{1}{Z} \, \left[ 1 + \frac{(q-1) \beta (E_{An} - \langle E_A \rangle)}{1 + (1-q)S_A} \right]^{-\frac{1}{q-1}}
\ea
The latter is exactly the Tsallis Type III distribution.  It is the one consistent with the fundamental hypothesis of equilibrium statistical mechanics.

\section{Single Particle Distributions}

Tsallis's definition of entropy, which leads to so-called ``Tsallis statistics", was not expected to describe a system of noninteracting particles or particles interacting with short-range forces \cite{Tsallis1998}.  Nevertheless, single particle distributions of the Tsallis type have been and are being used to describe momentum distributions in high energy collisions.  In this section I describe how those distributions come about.

Let's begin by considering one fermion degree of freedom with the Type III.  There is one quantum state which can be unoccupied with zero energy, or it can have one occupant with energy $\omega$.  The probabilities may be written as
\ba
p_0 &=& Z^{-1} \left[ 1 + (1-q) (S + \beta E ) \right]^{- \frac{1}{q-1}} \nonumber \\
p_1 &=& Z^{-1} \left[ 1 + (1-q) (S + \beta E - \beta \omega) \right]^{- \frac{1}{q-1}}
\label{p's}
\ea
The number operator $\hat{N}$ commutes with the Hamiltonian $H = \hat{N} \omega$.  Its average is
\be
\langle \hat{N} \rangle = \frac{p_1^q}{p_0^q + p_1^q}
\label{Nhatfermion}
\ee
and not $p_1/(p_0 + p_1)$ as one would have expected.  Explicitly
\be
\langle \hat{N} \rangle = \frac{1}{\left[ 1 + (q-1) \beta^* \omega \right]^{\frac{q}{q-1}} + 1}
\ee
and
\be
E \equiv \langle E \rangle = \frac{\omega}{\left[ 1 + (q-1) \beta^* \omega \right]^{\frac{q}{q-1}} + 1}
\ee
where
\be
\beta^* = \frac{\beta}{1 + (1-q) (S + \beta E )}
\ee
There is a self-consistency condition on these results.  To solve it, one could choose a value of $p_0$ between 0 and 1.  That then determines $p_1$, $S$ and $\langle E \rangle$ and hence $\beta$.  The classical Boltzmann limit is when $p_1 \ll p_0$ in which case
\be
\langle \hat{N} \rangle \rightarrow \left[ 1 + (q-1) \beta \omega \right]^{-\frac{q}{q-1}}
\label{classicalN}\
\ee
and $\beta^* \rightarrow \beta$.  This limit corresponds to $\beta \omega \gg 1$ and $q > 1$.

Next consider an arbitrarily large number of independent states with a set of quantum numbers and/or momenta labeled by $\alpha$.  The Hamiltonian is $H = \sum_{\alpha} \hat{N}_{\alpha} \omega_{\alpha}$ with number operators $\hat{N}_{\alpha}$.  The probabilities for these independent states factorize and are written as
\ba
p_{\alpha 0} &=& Z^{-1}_{\alpha} \left[ 1 + (1-q) (S_{\alpha} + \beta E_{\alpha} ) \right]^{- \frac{1}{q-1}} \nonumber \\
p_{\alpha 1} &=& Z^{-1}_{\alpha} \left[ 1 + (1-q) (S_{\alpha} + \beta E_{\alpha} - \beta \omega_{\alpha}) \right]^{- \frac{1}{q-1}}
\ea
with the averages for each state being 
\ba
N_{\alpha} &=& \frac{1}{\left[ 1 + (q-1) \beta^*_{\alpha} \omega_{\alpha} \right]^{\frac{q}{q-1}} + 1} \nonumber \\
E_{\alpha} &=& \frac{\omega_{\alpha}}{\left[ 1 + (q-1) \beta^*_{\alpha} \omega_{\alpha} \right]^{\frac{q}{q-1}} + 1}
\ea
and where
\be
\beta^*_{\alpha} = \frac{\beta}{1 + (1-q) (S_{\alpha} + \beta E_{\alpha} )}
\label{bstar}
\ee
The total energy is
\be
E = \sum_{\alpha} E_{\alpha}
\ee
and the total number of particles is
\be
N = \sum_{\alpha} N_{\alpha}
\ee
Note that there is a common temperature $T = 1/\beta$ but that $\beta^*_{\alpha}$ is state--dependent.  For a gas of particles 
\be
\omega_{\alpha} \rightarrow \omega(k)
\ee
where $k$ is the momentum and
\be
\sum_{\alpha} \rightarrow (2s+1) \int \frac{ d^3x d^3k}{(2\pi)^3}
\ee
with $s$ the spin of the fermions.

Any number of bosons may occupy a given quantum state.  This means that for bosons
\be
p_n = Z^{-1} \left[ 1 + (1-q) (S + \beta E - n \beta \omega) \right]^{- \frac{1}{q-1}}
\ee
The average of the number operator is
\be
\langle \hat{N} \rangle = \frac{\sum_{n=1}^{\infty} n p_n^q}{\sum_{m=0}^{\infty} p_m^q} =
\frac{\sum_n n \left[ 1 + (q-1) \beta^* \omega n \right]^{-\frac{q}{q-1}}}
{\sum_m \left[ 1 + (q-1) \beta^* \omega m \right]^{-\frac{q}{q-1}}}
\ee
where $\beta^*$ is calculated the same way as in (\ref{bstar}).  This expression cannot be evaluated in closed form for arbitrary $q$.  Several methods exist for evaluating it numerically \cite{Tsallis1998}.  One might have guessed that it was equal to
\bd
\frac{1}{\left[ 1 + (q-1) \beta^* \omega \right]^{\frac{q}{q-1}} - 1}
\ed
but it is not.  In the limit of classical Boltzmann statistics (small occupation probabilities) it reduces to (\ref{classicalN}), and in the limit $q \rightarrow 1$ it reduces to the usual Bose-Einstein distribution
\be
\langle \hat{N} \rangle = \frac{1}{e^{\beta \omega} - 1}
\ee
The extension to a gas of bosons follows in the same way as for fermions.

\section{Boltzmann and Fokker-Planck Equations}

The question naturally arises as to what sort of transport equation might give rise to a single particle Tsallis distribution.  In this section we describe a Boltzmann equation and a Fokker-Planck equation which would do so.

\subsection{Boltzmann Equation}

Consider the Boltzmann equation for the reaction $a+b \rightarrow c+d$ and its inverse.  It is
\ba
\frac{df_a}{dt} &=&  \int \frac{d^3p_b}{(2\pi)^3} \frac{d^3p_c}{(2\pi)^3} \frac{d^3p_d}{(2\pi)^3}
\Bigg\{ \frac{1}{1 + \delta_{cd}} W(c+d \rightarrow a+b) f_c f_d \left( 1+(-1)^{2s_a} f_a \right) \left( 1+(-1)^{2s_b} f_b \right) \nonumber \\
&-& \frac{1}{1 + \delta_{ab}} W(a+b \rightarrow c+d) f_a f_b \left( 1+(-1)^{2s_c} f_c \right) \left( 1+(-1)^{2s_d} f_d \right) \Bigg\}
\label{B}
\ea
This includes Pauli suppression for fermions and Bose enhancement for bosons in the final state.  The $s_i$ is the spin of particle $i$.  The coefficients take into account the possibility that the particles in the initial state are identical.  There is a gain term and a loss term.  Microscopic physics originating in quantum mechanics or quantum field theory says that
\be
\frac{1}{1 + \delta_{ab}} W(a+b \rightarrow c+d) = \frac{1}{1 + \delta_{cd}} W(c+d \rightarrow a+b)
\label{W}
\ee
The $W$'s are proportional to the square of a dimensional scattering amplitude $|{\cal M}|^2$ which in turn is proportional to the differential cross section $d\sigma/d\Omega$ in the center of momentum frame.  At tree level they are calculated from the vertices in the Lagrangian or Hamiltonian, while loop corrections also involve the propagators.  They do not depend on the distribution functions $f$.  In addition the $W$'s are proportional to an energy--momentum conserving $\delta$ function $\delta (p_a + p_b - p_c - p_d)$.  In perturbation theory, relevant to the Boltzmann equation, the equilibrium distributions are
\be
f = \frac{1}{\exp({\beta \omega}) - (-1)^{2s}}
\label{fequil}
\ee
In equilibrium $df_a/dt = 0$.  Using Eqs. (\ref{B}) and (\ref{W}) this leads to the condition $\omega_a + \omega_b = \omega_c + \omega_d$, in other words energy conservation.  This is also true in the limit of classical statistics.

On the other hand, consider using one of the Tsallis distributions.  We choose the limit of classical Boltzmann statistics because that limit is the same for Tsallis Types II and III.  It is 
\be
f = \left[ 1 + (q-1) \beta \omega \right]^{-\frac{q}{q-1}}
\label{fq}
\ee
In order that $df_a/dt = 0$ requires
\be
\omega_a + \omega_b + (q-1) \beta \omega_a \omega_b = \omega_c + \omega_d + (q-1) \beta \omega_c \omega_d
\ee
This is inconsistent with energy conservation and the $\delta$ function constraint unless $q = 1$ which is conventional statistical mechanics.  One can only imagine what happens when a collision involves $m$ particles in the initial state and $n$ particles in the final state.  Thus the aforementioned Tsallis distribution cannot be time independent.  To address this problem it has been {\it postulated} \cite{Lavagno2002} that the Boltzmann equation be modified so that the factor $f_a f_b$ (with classical statistics) be replaced with 
\be
h_q[f_a,f_b] \equiv \left( f_a^{1-q} + f_b^{1-q} -1 \right)^{\frac{1}{1-q}}
\ee
and that the equilibrium distribution function be\footnote{The difference in the power between (\ref{fq}) and (\ref{f1}) is whether one takes $f$ proportional to $p$, as in (\ref{p's}), or to $\langle \hat{N} \rangle$, as in (\ref{Nhatfermion}).  It should be the latter for consistency, but that does not affect the argument.}    
\be
f_{\rm eq} = \left[ 1 + (q-1) \beta \omega \right]^{\frac{1}{1-q}}
\label{f1}
\ee
By doing this it is easy to see that $\omega_a + \omega_b = \omega_c + \omega_d$.  There is a mention that the ``molecular chaos hypothesis" used to derive the original Boltzmann equation is not valid for Tsallis statistics but no proof or quantitative reasoning was given.  It would seem that this postulated modification is not consistent with the definition and physical interpretation of the $W$'s but is only a mathematical solution to the problem.

\subsection{Fokker-Planck Equation}

Consider the stochastic differential equation for the variable $u$
\be
\frac{du}{dt} = f(u) + g(u) \xi(t) + \eta(t)
\ee
where $\xi(t)$ and $\eta(t)$ are independent random variables.   They have white noise with averages
\be
\langle \xi(t) \rangle = \langle \eta(t) \rangle = 0
\ee
and correlation functions
\ba
\langle \xi(t) \xi(t') \rangle &=& 2M \delta (t - t') \nonumber \\
\langle \eta(t) \eta(t') \rangle &=& 2A \delta (t - t')
\ea
The $\xi(t)$ represents multiplicative noise and the $\eta(t)$ represents additive noise.  The special case where
\be
f = -\frac{c}{2} \frac{d g^2}{du}
\ee
and $g(0) = 0$ was studied in Ref. \cite{Anteneodo}.  See also Ref. \cite{Biro1}.  The steady state probability distribution to the resulting Fokker-Planck equation has the Tsallis form
\be
P(u) \sim \left[ 1 + (q-1) \beta g^2(u) \right]^{- \frac{1}{q-1}}
\ee
where $q = (c+3M)/(c+M)$ and $\beta = (c+M)/2A$. We consider two possibilities that come closest to describing the single particle distribution.  We choose 
$u = k$, the momentum, and $\omega = \sqrt{k^2 + m^2}$, with $m$ the particle mass.  We also define $c = 2 m \nu$. 

The first possibility is $g = p/\sqrt{2m}$.  This results in
\be
\frac{dp}{dt} = - \nu p + \frac{p}{\sqrt{2m}} \xi(t) + \eta(t)
\ee
and
\be
P(u) \sim \left[ 1 + (q-1) \beta \frac{p^2}{2m} \right]^{- \frac{1}{q-1}}
\ee
This is the nonrelativistic limit.  The second possibility is $g = \sqrt{\omega - m}$.  This results in
\be
\frac{dp}{dt} = - m \nu \frac{p}{\omega} + \sqrt{\omega - m} \, \xi(t) + \eta(t)
\ee
and
\be
P(u) \sim \left[ 1 + (q-1) \beta (\omega - m) \right]^{- \frac{1}{q-1}}
\ee

Additive noise is normal for particles moving in a heat bath.  Multiplicative noise is expected in hydrodynamic descriptions of high energy nucleus-nucleus collisions \cite{Youngsolo}.  The problem is that the results are highly sensitive to the choice of parameters in the Fokker-Planck equation.  Typically they must be fine-tuned to 4 or 5 significant figures, and they depend on the collision system, beam energy, and hadron species, as we shall see in the next section.

\section{High Energy Collision Phenomenology}

Tsallis-like or Tsallis-inspired distributions have been used many times by experimental collaborations to parametrize particle spectra in high-energy proton-proton collisions.  For a few examples see \cite{STAR2007,PHENIX2011,ALICE2011,CMS2012}.  Generally they are of the form
\be
\frac{dN}{dy d^2p_T} \propto \frac{dN}{dy} \left( 1 + \frac{m_T - m}{nT} \right)^{-n}
\label{datafit}
\ee
where $y$ is the rapidity, $m_T = \sqrt{m^2 + p_T^2}$ is the transverse mass, and $p_T$ is the transverse momentum.  This distribution is approximately 
$\exp(-m_T/T)$ at low $p_T$ and $p_T^{-n}$ at high $p_T$.  If one makes the identification $n = 1/(q-1)$ then one would associate $T$ with the temperature, whereas if one makes the identification $n = q/(q-1)$ then one would associate $qT$ with the temperature.  According to Sec. IV it really should be the latter.  In the following I will use the same notation as in the original papers.

A compilation of charged hadron distributions from \cite{Wong2015} is shown in Fig. 1.  The fits are to the function
\bd
A \left( 1 + (q-1) \beta m_T \right)^{-1/(q-1)}
\ed
\begin{figure}[h]
\center{\includegraphics[width=0.93\linewidth]{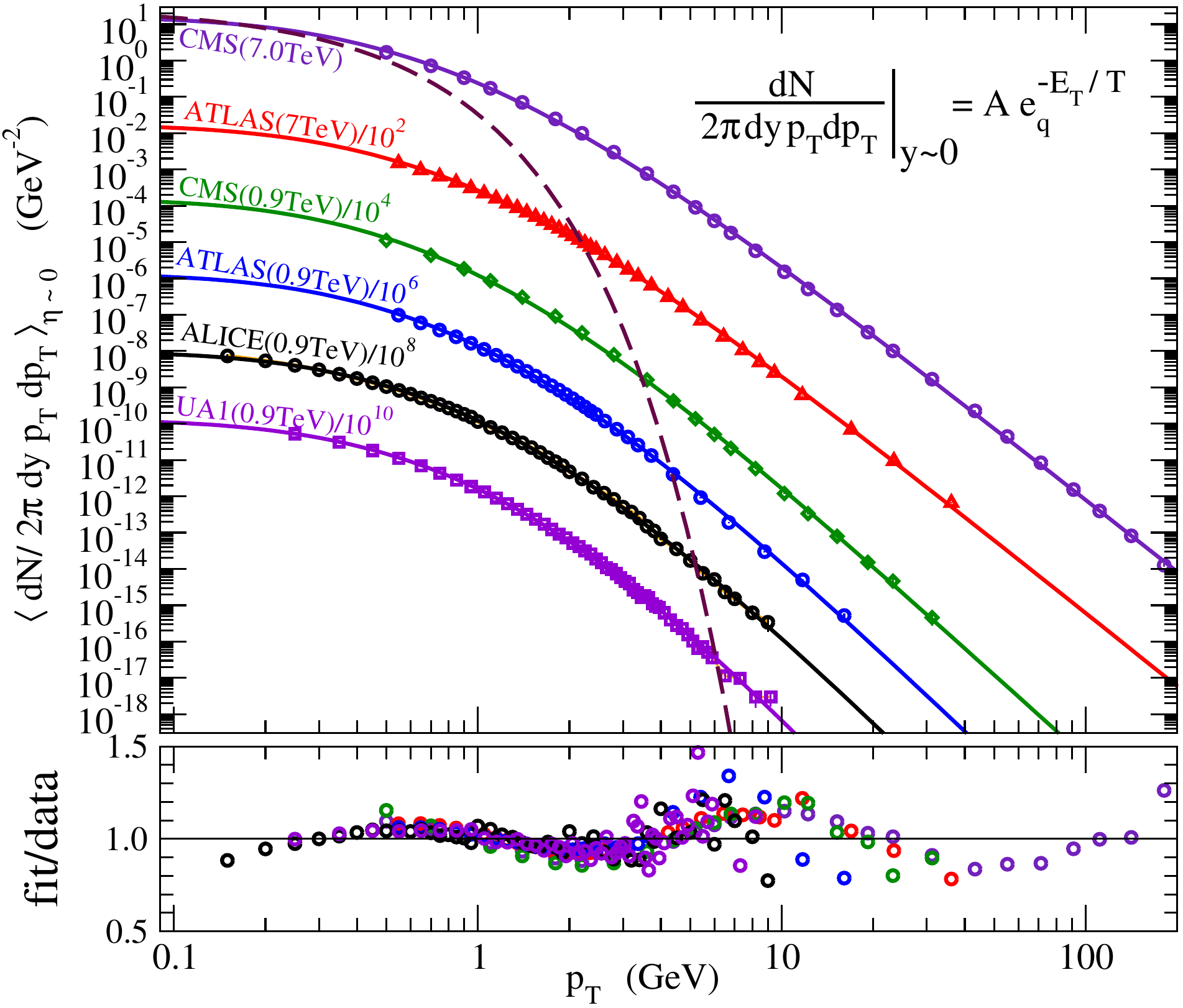}}
\caption{Three-parameter fits to $pp$ collisions with $A \left( 1 + (q-1) \beta m_T \right)^{-1/(q-1)}$.  From \cite{Wong2015}.}
\end{figure}
Different experiments have somewhat different experimental cuts, such as the rapidity window or minimum multiplicity, so the fit parameters can be slightly different at the same beam energy.  It is amazing that such a simple function could represent data over up to 14 orders of magnitude.  However, it is necessary to adjust $q$ to 4 significant figures, ranging from 1.109 for ATLAS at 0.9 TeV to 1.151 for CMS at 7 TeV.  Although this may seem to be a rather narrow range, they translate to very different powers of $1/(q-1)$ = 9.21 and 6.60, respectively.   

Comparisons to data taken at RHIC for the most central d + Au, Cu + Cu, Au + Au and at the LHC for p + Pb and Pb + Pb show that the $n$ in Eq. (\ref{datafit}) depends on the colliding systems, beam energy, and particle species \cite{Zheng2015}.  Comparisons to data taken at RHIC and at LHC for proton-proton, proton-nucleus, and nucleus-nucleus collisions show that it also depends on the multiplicity \cite{Biro2}.

The underlying physics in these collisions is QCD.  Low transverse momentum hadrons are produced with the normal kinetic and chemical equilibrium distributions boosted by collective hydrodynamic flow.  They essentially exhibit exponential decay in energy.  High transverse momentum hadrons exhibit power-law decay because of the asymptotic freedom of QCD.  The parton model for scattering of point particles 
\be
\frac{d\sigma}{dy d^2p_T} \sim p_T^{-n}
\ee
predicts $n=4$ simply from dimensional analysis and therefore $q=4/3$ if one identifies $n$ with $q/(q-1)$.  However, scale breaking and realistic parton distribution functions in the projectile and target increase $n$ significantly and so numerically $q$ turns out to be much closer to 1 for hadrons, although $n \sim$ 4.5 to 5.5 for jets \cite{Wong2015,Wong2013}.  Basically the Tsallis distributions have an extra parameter, $q$, which controls the transition from exponential to power-law behavior.  It is an efficient parametrization of data and/or perturbative QCD based parton models, but should not be considered a fundamental constant.  Quoting a value for $n$ rather than for $q$ is actually more revealing of the underlying physics.  

\section{Conclusion}

Tsallis and collaborators proposed three types of nonextensive statistical mechanics.  They all assumed an entropy of the form (\ref{Tentropy}) but differed in how averages of conserved quantities were defined.  It turns out that only Type III is consistent with the fundamental hypothesis of equilibrium statistical mechanics.  Everything else follows without ambiguity.  Notably there are nontrivial self-consistency conditions to be solved.  These self-consistency conditions must be solved mode by mode for single particle distribution functions.  In the limit of small occupation probabilities (classical Boltzmann statistics) these conditions simplify and basically are satisfied by the normalization of probabilities.  With Type III energies are additive.  However, the single particle distribution functions are not time independent solutions of Boltzmann's equation.  They are time independent solutions of a modified Boltzmann equation with fractional exponents of the distributions appearing, but this does not appear to be physically consistent with the microscopic derivation of scattering amplitudes.  Tsallis-inspired single particle distribution functions provide an efficient parametrization of high energy collision data but have not been derived from QCD.  The parameter $q$ encapsulates a lot of complicated physics in these collisions, but nonextensivity is not part of that physics.

\section*{Acknowledgement}
I am very grateful to P. Arnold, T. Biró, S. Rudaz, C. Tsallis, and C.-Y. Wong for constructive comments on the manuscript.  This work was supported by the U.S. Department of Energy Grant DE-FG02-87ER40328.

\end{document}